\documentstyle{article}
\newcounter{bla}
\newenvironment{refnummer}{%
\list{[\arabic{bla}]}%
{\usecounter{bla}%
 \setlength{\itemindent}{0pt}%
 \setlength{\topsep}{0pt}%
 \setlength{\itemsep}{0pt}%
 \setlength{\labelsep}{2pt}%
 \setlength{\listparindent}{0pt}%
 \settowidth{\labelwidth}{[9]}%
 \setlength{\leftmargin}{\labelwidth}%
 \addtolength{\leftmargin}{\labelsep}%
 \setlength{\rightmargin}{0pt}}}
 {\endlist}

\makeatletter

\def\AmS{{\protect\the\textfont\tw@
  A\kern-.1667em\lower.5ex\hbox{M}\kern-.125emS}}
\makeatother

\begin{document}

\title{The library of subroutines for calculating standard quantities in atomic structure theory\\}

\author{ Gediminas Gaigalas\\
        {\em State Institute of Theoretical Physics and Astronomy,}\\
        {\em A. Go\v{s}tauto 12, Vilnius 2600, LITHUANIA}}

\maketitle

\begin{abstract}
This library (collection of subroutines) is presented for calculating standard quantities 
in the decomposition of
many--electron matrix elements in atomic structure theory. These quantities include the
coefficients of fractional parentage, the reduced coefficients of fractional parentage 
as well as reduced and completely reduced matrix elements for several operators.
So the library is assigned for any computational scheme.
The software is an implementation of a methodology based on the second
quantization in coupled tensorial form, the angular momentum theory in
3 spaces (orbital, spin and quasispin), and the graphical technique of
angular momentum. This implementation
extends applications in atomic theory capabilities to partially filled $f$-- shells
and has lead to faster execution of angular integration codes.
\end{abstract}

\vspace{6cm} {\em PACS:} 31.15.Ne, 31.25.-v, 32.10.-f, 32.30.-r

\vspace{0.3cm}

\newpage

{\bf PROGRAM SUMMARY} \\

\begin{small}


\noindent {\em Title of program} {\em :} SQ \\[10pt]
{\em Catalogue identifier:} \\[10pt]
{\em Program obtainable from:} State Institute of Theoretical Physic and Astronomy,
A. Go\v stauto 12, Vilnius, 2600, LITHUANIA.~~ E-mail: gaigalas@itpa.lt \\[10pt]
{\em Computer for which the library is designed and others on
which it has been tested:}\\
{\em Computers:} Pentium--based PCs\\[10pt]
{\em Installations:} Institute of Theoretical Physics and Astronomy,
A. Go\v stauto 12, Vilnius, 2600, LITHUANIA \\[10pt]
{\em Operating systems or monitors under which the new version has been
tested:} Sun UNIX, LINUX 2.2.3 \\[10pt]
{\em Programming language used in the new version:} FORTRAN 77 \\[10pt]
{\em Memory required to execute with typical data:} 6000K Bytes  \\[10pt]
{\em Peripherals used:} terminal, disk\\[10pt]
{\em No. of bits in a word:} 32\\[10pt]
{\em No. of processors used:} 1\\[10pt]
{\em Has the code been vectorised or parallelized?:} no \\[10pt]
{\em No. of bytes in distributed program, including test data, etc.:}
285 000 bytes \\[10pt]
{\em Distribution format:} compressed tar file \\[10pt]
{\em CPC Program Library subprograms used:} none \\[10pt]
{\em Additional keywords :} atomic structure, Clebsch--Gordan coefficient, 
configuration interaction,
Reduced coefficients of fractional parentage, irreducible tensors,
angular momentum theory in three spaces (orbit, spin and quasispin),
second quantization in the coupled tensorial form, recoupling coefficients,
complex atom, correlation,
wave functions, bound states, $LS$ coupling, $f$-- shell, matrix elements, 
$3nj$--coefficients.
\\[10pt]
{\em Nature of physical problem}\\
Accurate theoretical determination of atomic energy levels, orbitals and radiative
transition data requires the calculation of matrix elements of physical operators 
accounting for relativistic and correlation effects (see
the multiconfiguration Hartree--Fock method~[1], for example). 
The spin--angular integration of these matrix elements is typically based on
standard quantities like the matrix elements of the unit tensor, the (reduced)
coefficients of fractional parentage as well as a number of other reduced 
matrix elements concerning various products of electron creation and 
annihilation operators~[2].
These quantities arise very frequently both in configuration interaction approaches
and the derivation of perturbation expansions for many--particle system using 
symmetry--adapted configuration state functions.\\[10pt]
{\em Method of solution} \\
This program is created involving the
angular methodology of [2--7]. It
has been extended to include partially filled $f$-- subshells
in wavefunction expansions.
The classification of terms is  identical to that described in [6].
\\[10pt]
{\em Restrictions on the complexity of the problem}\\
For $LS$--coupling subshells states, the library provides coefficients and matrix
elements for all subshells (nl) with $l=0$, $1$, $2$ and $3$, and $l^{2}$ for $l \ge 3$.
\\[10pt]
{\em Unusual features of the program }\\
The library can be used as an "electronic tables" of standard quantities for
evaluating general matrix elements for $LS$--coupled functions.
\\[10pt]
{\em References}

\begin{refnummer}
\item C. Froese Fischer, T. Brage and P. J\" onsson,
   {\sl Computational Atomic Structure. An MCHF Approach}
   (Institute of Physics Publishing, Bristol/Philadelphia, 1997).

\item Z. Rudzikas, {\sl Theoretical Atomic Spectroscopy}
   (Cambridge Univ. Press, Cambridge, 1997).

\item G. Gaigalas and Z. Rudzikas,
J. Phys. B: At. Mol. Phys. 29 (1996) 3303.

\item G. Gaigalas, Z. Rudzikas and C. Froese Fischer,
J. Phys. B: At. Mol. Phys. 30 (1997) 3747.

\item G. Gaigalas, A. Bernotas, Z. Rudzikas and C. Froese Fischer,
Physica Scripta 57 (1998) 207.

\item G. Gaigalas, Z. Rudzikas and C. Froese Fischer,
Atomic Data and Nuclear Data Tables 70 (1998) 1.

\item G. Gaigalas, Lithuanian Journal of Physics 39 (1999) 80.
\end{refnummer}

\end{small}


\clearpage

\section*{LONG WRITE--UP}

\section{Introduction}

In order to obtain accurate values of atomic quantities
it is necessary to account for relativistic and correlation effects
(see for example \cite{R,Fbook}).
Relativistic effects may be taken into account as Breit--Pauli
corrections or in a fully relativistic approach. In both cases for
complex atoms and ions, a considerable part of the effort must be
devoted to integrations over spin--angular variables,
occurring in the matrix elements of the operators under consideration.

\medskip

Many existing codes for integrating
are based on a scheme first proposed by Fano~\cite{Fano}.
This approach is based on the coefficients of fractional parentage (CFP)
and then the integrations over spin--angular variables constitute a
considerable part of the computation, especially when atoms with many open
shells are treated, and the operators are non--trivial \cite{GlassH}. 
Over the last decade,
an efficient approach for finding matrix elements of any
one-- and two--particle atomic operator between complex configurations has been
developed (see papers Gaigalas~{\it et al}
\cite{method1,method2,method3,method5,method7}).
It is free of the shortcomings of previous approaches
(see Gaigalas \cite{method6}).
This approach is based on a second quantization \cite{Judd-s} and uses a coupled tensorial 
form for the electron creation and annihilation operators \cite{RK}. It also applies the 
theory of angular momentum \cite{JB,JS}
to three different spaces, i.e. the space of orbital angular momentum
$l$, spin space $s$ and the quasispin space $q$ \cite{R}. The basic quantities of this new 
approach are 
the reduced coefficients of fractional parentage (RCFP) and the completely reduced matrix
elements of the $W^{(k_q k_l k_s)}$ operator.

\medskip

Obviously, each computational scheme is based on a set of standard quantities to
decompose the many--electron matrix elements. These quantities are either CFP, RCFP, the
reduced matrix elements of the unit tensors $U^{(k)}$ and $V^{(k1)}$, the completely 
reduced matrix elements $W^{(k_q k_l k_s)}$, the completely reduced matrix elements
of some tensorial products of second quantization operators, depending on the approach
\cite{R}.
In this paper, we will present the library SQ (standard quantities). This library is 
collection of subroutines for the calculation 
of abovementioned standard quantities. The same standard quantities 
arise very frequently in effective Hamiltonian of
perturbation theory or effective operator whose matrix elements between the 
non--relativistic $LS$ coupling states are equal to the matrix elements of the full
electronic hamiltonian, between the corresponding  $jj$ coupled relativistic states (see
Rudzikas~\cite{R}), too. So the library not only supports large--scale computations of
open--shell atoms using multiconfiguration Hartree--Fock or configuration interaction
approaches, but may even help to develope codes for calculating the angular parts of
effective operators from many--body perturbation theory and orthogonal operators
or for evaluating relativistic
hamiltonian in $LS$--coupling as well as for various versions of semi--empirical methods.
The code also is intented for approaches and/or calculations presented in 
\cite{Karazija-book,Karazija-book-b,GaigalasKKMV:94,BogGMR:97,GobGM:98}. 
Some very accurate calculations was performed using this library (see for example
\cite{BogKKKG:97,FroeseSGG:98,FroeseYG:95,FroeseGG:97}), too.

\medskip

The theoretical background of this program will be presented in the section 2. It 
includes a brief outline
of the quasispin concept, the defininitions of the RCFP and the reduced matrix elements of 
$W^{(k_q k_l k_s)}$. The library is presented in section 3.

\section{Theoretical bacground}

The library is based on the irreducible tensorial form of the second quantization 
operators and on a quasispin technique. In this section we briefly describe
this approach.

\medskip

In the quasispin representation, for a wave function of the shell of
equivalent electrons $|nl^N\alpha LS)$ a label $Q$ -- quasispin momentum of
the shell -- is introduced, which is related to the seniority quantum number
$\nu $, namely, $Q=\left( 2l+1-\nu \right) /2$, and its projection,
$M_Q=\left( N-2l-1\right) /2$. Here $\alpha $ denotes all additional quantum
numbers needed for the one--to--one classification of the energy levels. Then,
we can rewrite the wavefunction of equivalent electrons as
\begin{eqnarray}
\label{eq:the-a}
   |nl^N\alpha QLS \: M_Q) .
\end{eqnarray}
Making use of the Wigner--Eckart theorem in quasispin space of a shell $l^N$,
\begin{eqnarray}
\label{eq:gghhaa}
\lefteqn{
   \left( l^N\;\alpha QLSM_Q||T_{m_q}^{\left( k_qk_lk_s\right) }||l^{N^{\prime}}\;
   \alpha ^{\prime
   }Q^{\prime }L^{\prime }S^{\prime }M_Q^{\prime }\right) }
   \nonumber  \\[1ex]
   & & =
   \left( -1\right)
   ^{2k_q}\left[ Q\right] ^{-1/2}
   \left[
   \begin{array}{ccc}
   Q^{\prime } & k_q & Q \\
   M_Q^{\prime } & m_q & M_Q
   \end{array}
   \right]
   \left( l\;\alpha QLS|||T^{\left( k_qk_lk_s\right) }|||l\;\alpha ^{\prime
   }Q^{\prime }L^{\prime }S^{\prime }\right)
\end{eqnarray}
it is possible to define the notions of a completely reduced matrix element
$\left( l\;\alpha QLS|||T^{\left( k_qk_lk_s\right) }|||l\;\alpha ^{\prime
}Q^{\prime }L^{\prime }S^{\prime }\right) $. In (\ref{eq:gghhaa}) 
$T_{m_q}^{\left( k_qk_lk_s\right) }$
is any tensor with rank $k_q$ and its projection $m_q$ in quasispin space and
on the right--hand side of this equation only the Clebsch--Gordan coefficient
$\left[
\begin{array}{ccc}
Q^{\prime } & k_q & Q \\
M_Q^{\prime } & m_q & M_Q
\end{array}
\right] $ depends on the number $N$ of equivalent electrons.

\medskip

The electron creation $ a_{m_{l}\:m_{s}}^{\left(l\:s \right)} $ and annihilation 
$ a_{-m_{l}\: -m_{s}}^{\left(l \: s \right) \dagger} $ operators play a key role in 
the theory of second quantization and atomic structure \cite{Judd-s}. 
Using the quasispin concept, the 
operators $ a_{m_{l}\:m_{s}}^{\left(l\:s \right)} $ and 
$ \stackrel{\sim}{a}_{m_{l}\:m_{s}}^{\left(l\:s \right)}=\left( -1\right)^{l+s-m_{l}-m_{s}}
a_{-m_{l}\: -m_{s}}^{\left(l \: s \right) \dagger} $ also form components of an 
irreducible tensor of rank $q=\frac 12$ in $Q$--space, i.e.\
\begin{equation}
   \label{eq:second-a}
   a_{\, m_q \: m_l \: m_s}^{\left( q \: l \: s\right) }=\left\{
   \begin{array}{ll}
   a_{\, m_l \: m_s}^{( l \: s )}  & \hspace*{1cm} \mbox{ for } m_q  = \frac 12, \\[0.3cm]
   \stackrel{\sim }{a}_{\, m_l \: m_s}^{( l \: s ) } 
                       & \hspace*{1cm} \mbox{ for } m_q = -\frac 12.
   \end{array}
   \right.  
\end{equation}
Compared with the electron creation and annihilation operators above, 
the operators $a_{\, m_q \: m_l \: m_s }^{\left( q \: l \: s\right) }$ also act in an 
additional quasispin space like a tensor component with
rank $q$ and a projection $m_q=\pm \frac 12$.
There is the following relation known between the reduced matrix element 
of a creation operator and the CFP \cite{La-Ma}:
\begin{eqnarray}
   \label{eq:cfp-amat}
   \left( l^N \;\alpha QLS||a^{\left( ls \right) }||l^{N-1} \;\alpha ^{\prime
   }Q^{\prime }L^{\prime }S^{\prime }\right)
   & = &
   \left( -1\right) ^{N}
   \sqrt{N\left[ J,L,S\right]}
   \left( l^N\;\alpha QLS||l^{N-1}\;
   \left( \alpha ^{\prime }Q^{\prime}L^{\prime }S^{\prime }
   \right) \; l
   \right) ,
\end{eqnarray}
where $\left[ L,S \right] \equiv \left( 2L +1\right)\left( 2S +1\right)$.
Eqs.\ (\ref{eq:cfp-amat}) and (\ref{eq:gghhaa}) can be used to define the relation between 
the CFP and its reduced counterpart in $Q$--space. Introducing the $z-$projection, 
$M_Q$, of the quasispin, this relation is given by \cite{R}

\begin{equation}
\label{eq:gi}
\begin{array}[b]{c}
\left( l^N\;\alpha QLS||l^{N-1}\;\left( \alpha ^{\prime }Q^{\prime
}L^{\prime }S^{\prime }\right) l\right) =\left( -1\right) ^{N-1}\left(
N\left[ Q,L,S\right] \right) ^{-1/2}\times \\
\times \left[
\begin{array}{ccc}
Q^{\prime } & 1/2 & Q \\
M_Q^{\prime } & 1/2 & M_Q
\end{array}
\right] \left( l\;\alpha QLS|||a^{\left( qls\right) }|||l\;\alpha ^{\prime
}Q^{\prime }L^{\prime }S^{\prime }\right) .
\end{array}
\end{equation}

Tables of numerical values of $\left( l\;\alpha QLS|||a^{\left( qls\right)
}|||l\;\alpha Q^{\prime }L^{\prime }S^{\prime }\right) $ are presented in
Rudzikas and Kaniauskas \cite{RK} when $l=0,1,2$. For the tensorial product of
two one--electron operators, the submatrix element equals

\begin{equation}
\label{eq:gj}
\begin{array}[b]{c}
\left( nl^N\;\alpha QLS||\left[ a_{m_{q1}}^{\left( q\lambda \right) }\times
a_{m_{q2}}^{\left( q\lambda \right) }\right] ^{\left( k_1k_2\right)
}||nl^{N^{\prime }}\;\alpha ^{\prime }Q^{\prime }L^{\prime }S^{\prime
}\right) = \\
=\displaystyle {\sum_{\epsilon ,m_\epsilon }}\left[ Q\right] ^{-1/2}\left[
\begin{array}{ccc}
q & q & \epsilon \\
m_{q1} & m_{q2} & m_\epsilon
\end{array}
\right] \left[
\begin{array}{ccc}
Q^{\prime } & \epsilon & Q \\
M_Q^{\prime } & m_\epsilon & M_Q
\end{array}
\right] \times \\
\times \left( nl\;\alpha QLS|||W^{\left( \epsilon k_1k_2\right)
}|||nl\;\alpha ^{\prime }Q^{\prime }L^{\prime }S^{\prime }\right) .
\end{array}
\end{equation}

On the right--hand side of equations (\ref{eq:gi}) and (\ref{eq:gj}) only the
Clebsch--Gordan coefficient $\left[
\begin{array}{ccc}
Q^{\prime } & \epsilon & Q \\
M_Q^{\prime } & m_\epsilon & M_Q
\end{array}
\right] $ depends on the number $N$ of equivalent electrons.

$\left( nl\;\alpha QLS|||W^{\left( \epsilon k_1k_2\right) }|||nl\;\alpha
^{\prime }Q^{\prime }L^{\prime }S^{\prime }\right) $ denotes reduced in
quasispin space submatrix element (completely reduced matrix element) of the
triple tensor $W^{\left( \epsilon k_1k_2\right) }
=\left[
a^{\left( qls\right) }\times a^{\left( qls\right) }\right] ^{\left( \epsilon
k_1k_2\right) }$. It is related to the RCFP in a following way:

\begin{equation}
\label{eq:gk}
\begin{array}[b]{c}
\left( nl\;\alpha QLS|||W^{\left( \epsilon k_1k_2\right) }|||nl\;\alpha
^{\prime }Q^{\prime }L^{\prime }S^{\prime }\right) = \\
=\left( -1\right) ^{Q+L+S+Q^{\prime }+L^{\prime }+S^{\prime }+\epsilon
+k_1+k_2}\left[ \epsilon ,k_1,k_2\right] ^{1/2}\times \\
\times
\displaystyle {\sum_{\alpha ^{\prime \prime }Q^{\prime \prime }L^{\prime
\prime }S^{\prime \prime }}}\left( l\;\alpha QLS|||a^{\left( qls\right)
}|||l\;\alpha ^{\prime \prime }Q^{\prime \prime }L^{\prime \prime }S^{\prime
\prime }\right) \times \\ \times \left( l\;\alpha ^{\prime \prime }Q^{\prime
\prime }L^{\prime \prime }S^{\prime \prime }|||a^{\left( qls\right)
}|||l\;\alpha ^{\prime }Q^{\prime }L^{\prime }S^{\prime }\right) \times \\
\times \left\{
\begin{array}{ccc}
q & q & \epsilon \\
Q^{\prime } & Q & Q^{\prime \prime }
\end{array}
\right\} \left\{
\begin{array}{ccc}
l & l & k_1 \\
L^{\prime } & L & L^{\prime \prime }
\end{array}
\right\} \left\{
\begin{array}{ccc}
s & s & k_2 \\
S^{\prime } & S & S^{\prime \prime }
\end{array}
\right\} .
\end{array}
\end{equation}

So, by applying the quasispin method for calculating the matrix elements of
any operator, we can use the RCFP or the completely reduced matrix elements
of $W^{\left( \epsilon k_1k_2\right) }\left(
nl,nl\right) $, which are independent of the occupation number of the shell.
The main advantage of this approach is that the standard
data tables in such a case will be much smaller in comparison with tables of
the usual coefficients and, therefore, many summations will be less
time--consuming. Also one can see that in such an approach the 
completely reduced matrix elements of standard tensors and RCFP
actually can be treated in a uniform way as they all are the completely
reduced matrix elements of the second quantization operators. Hence, all
methodology of calculation of matrix elements will be much more universal in
comparison with the traditional one.

\section{Description of the Library}

\subsection{SAI\_SQLS1}
\label{sec-sls}

The section SQLS1, (standard quantities in $LS$ coupling, part one)
is a collection of utilities for calculation of standard quantities as:
\begin{itemize}
\item Clebsh--Gordan coefficients of type
$\left[
\begin{array}{ccc}
Q   & \frac{1}{2} & C \\
m_Q & m_S & m_C
\end{array}
\right]$,
$\left[
\begin{array}{ccc}
Q   & 1 & C \\
m_Q & 0 & m_C
\end{array}
\right]$ and
$\left[
\begin{array}{ccc}
Q   & 1 & C \\
m_Q & 1 & m_C
\end{array}
\right]$.
\item The $6j-$ and $9j-$ coefficients.
\item The RCFP $(\ l\ QLS\ |||\ a^{(qls)}\ |||\ l\ Q'L'S'\ )$.
\item Reduced matrix element
$(\ l\ QLS\ |||\ [~a^{(qls)} \times a^{(qls)}~]^{(k_{1}k_{2}k_{3})}\ |||\ l\ Q'L'S'\ )$.
\item The matrix elements of type
$(\ l^{N}\ QLS\ ||\ [a^{(qls)}_{m_{q1}} \times a^{(qls)}_{m_{q2}}]^{(k_{l}k_{s})} ||\ 
l^{N'}\ Q'L'S'\ )$,

$(\ l^{N}\ QLS\ ||\ [ a^{(qls)}_{m_{q1}} \times [\ a^{(qls)}_{m_{q2}} \times 
a^{(qls)}_{m_{q3}}\ ]^{(k_{1}k_{2})} ]^{(k_{l}k_{s})}\ ||\ l^{N'}\ Q'L'S'\ )$,

$(\ l^{N}\ QLS\ ||\ [ [\ a^{(qls)}_{m_{q1}} \times a^{(qls)}_{m_{q2}}\ ]^{(k_{1}k_{2})} 
\times a^{(qls)}_{m_{q3}}\ ]^{(k_{l}k_{s})}\ ||\ l^{N'}\ Q'L'S'\ )$,

$(\ l^{N}\ QLS\ ||\ [\ [\ a^{(qls)}_{m_{q1}} \times a^{(qls)}_{m_{q2}}\ ]^{(k_{l}k_{s})} 
\times [\ a^{(qls)}_{m_{q3}} \times a^{(qls)}_{m_{q4}}\ ]^{(k_{l}k_{s})} ]^{(00)}\ ||\ 
l^{N'}\ QLS\ )$
and

$(\ l^{N}\ QLS\ ||\ [\ [\ a^{(qls)}_{m_{q1}} \times a^{(qls)}_{m_{q2}}\ ]^{(k_{1}k_{2})} 
\times [\ a^{(qls)}_{m_{q3}} \times a^{(qls)}_{m_{q4}}\ ]^{(k_{3}k_{4})} ]^{(k_{l}k_{s})}\ 
||\ l^{N'}\ QLS\ )$.

\end{itemize}

The subroutines presented in this section are independent and may be useful for other programs.
67 subroutines are contained in this library. Was created the 
library \cite{GaigalasM:86} on the low--powered computers BESM--6 of the 1980s
for some similar quantities for the $s$--, $p$-- and $d$-- shells. 
It was written in Fortran~4 and had 17 subroutines. It was based on the approach 
\cite{MerkelisG:85} and assign for calculation submatrix elements of the nonrelativistic 
effective Hamiultonian of atom in the first two orders of stationary perturbation theory 
in the case of configuration of two open--shells $n_{1} l^{N_1}_{1} \; n_{2} l^{N_2}_{2}$.
The present library is more general and has more future because it is useful for different
approaches with any number of open shells with $l=0$, $1$, $2$ and $3$, and $l^{2}$ for 
$l \ge 3$.

\subsubsection{Common Blocks}

Most of the subroutines use common block GLCONS from
MCHF atomic structure package \cite{Fbook,F}.
The parameters contained in this block are defined in BLOCK DATA GLCONS.

\medskip

The Common blocks /MT/ and /MT67/ are important, too. The first one has the
array MT(40), which contains all the term characteristics of $s$--, $p$--
and $d$-- shells, that are needed while calculating matrix elements in the quasispin
formalism (see Table \ref{ta-a}). An element of the MT array is indicated in the
column $No$ of Table \ref{ta-a}, and the $Term$ column indicates the
characteristics of a term contained in that element of MT array. In other
words, for all the $s$--, $p$-- and $d$-- shells, the terms are numbered from
1 to 40. The terms are marked as $_{~~~~~~\nu}^{(2S+1)}L^{(2Q+1)}$
in the table.

\medskip

All the terms for $f$-- shell (see Table \ref{ta-b}) are similarly placed in
the Common block /MT67/. The $No$ column of Table~\ref{ta-b} indicates the
term number. The $TERM$ column has $^{(2S+1)}L^{Nr}$ in it, the $2Q$
column has the quasispin momentum $Q$ multiplied by two.
See \cite{method5,GFa} for the details of the classification of
$f$-- shell terms.

\medskip

The Common block /MT/ is defined in BLOCK DATA TERMLS. Meanwhile the /MT67/
is defined in BLOCK DATA TRMF from the library {\bf SAI\_SQLS2}
(see below). BLOCK DATA TERMLS defines COMMON /SKMT2/, which contains
term characteristics for special cases, i.e. for shells $l=3-9$ and shell
occupation numbers $N = 1,2$.

\begin{table}
\begin{center}
\caption{Allowed couplings of $l^{N}$ states for $l = 0 - 2$. The subshell
quasispin angular momentum $Q$, spin angular momentum $S$, the subshell
angular momentum $L$ and seniority quantum number $\nu$
are denoted $_{~~~~~~\nu}^{(2S+1)}L^{(2Q+1)}$.}
\label{ta-a}
\begin{tabular}{rcc|rcc|rcc|rcc} \hline
No & Term &&  No & Term &&  No & Term &&
No & Term &  \\ \hline
\multicolumn{2}{l}{\sl subshell $s$} & &
\multicolumn{2}{l}{\sl subshell $d$} & &
19.& $^{2}_{5}F^{1}$  &&
30.& $^{3}_{4}D^{2}$  &\\
1. & $^{2}_{1}S^{1}$  &&
9. & $^{6}_{5}S^{1}$  &&
20.& $^{2}_{3}G^{3}$  &&
31.& $^{1}_{2}D^{4}$  &\\
2. & $^{1}_{0}S^{2}$  &&
10.& $^{2}_{5}S^{1}$  &&
21.& $^{4}_{5}G^{1}$  &&
32.& $^{1}_{4}D^{2}$  &\\
\multicolumn{2}{l}{\sl subshell $p$} &  &
11.& $^{4}_{3}P^{3}$  &&
22.& $^{2}_{5}G^{1}$  &&
33.& $^{3}_{2}F^{4}$  &\\
3. & $^{4}_{3}S^{1}$  &&
12.& $^{2}_{3}P^{3}$  &&
23.& $^{2}_{3}H^{3}$  &&
34.& $^{3}_{4}F^{2}$  &\\
4. & $^{2}_{1}P^{3}$  &&
13.& $^{2}_{1}D^{5}$  &&
24.& $^{2}_{5}I^{1}$  &&
35.& $^{1}_{4}F^{2}$  &\\
5. & $^{2}_{3}D^{1}$  &&
14.& $^{2}_{3}D^{3}$  &&
25.& $^{1}_{0}S^{6}$  &&
36.& $^{3}_{4}G^{2}$  &\\
6. & $^{1}_{0}S^{4}$  &&
15.& $^{4}_{5}D^{1}$  &&
26.& $^{1}_{4}S^{2}$  &&
37.& $^{1}_{2}G^{4}$  &\\
7. & $^{3}_{2}P^{2}$  &&
16.& $^{2}_{5}D^{1}$  &&
27.& $^{3}_{2}P^{4}$  &&
38.& $^{1}_{4}G^{2}$  &\\
8. & $^{1}_{2}D^{2}$  &&
17.& $^{4}_{3}F^{3}$  &&
28.& $^{3}_{4}P^{2}$  &&
39.& $^{3}_{4}H^{2}$  &\\
   &                  &&
18.& $^{2}_{3}F^{3}$  &&
29.& $^{5}_{4}D^{2}$  &&
40.& $^{1}_{4}I^{2}$  &\\ \hline
\end {tabular}
\end{center}
\end{table}

\begin{table}
\begin{center}
\caption{Allowed couplings of $f^{N}$. The subshell spin angular momentum $S$,
number $Nr$ and the subshell angular momentum $L$ are denoted $^{(2S+1)}L^{Nr}$.
The quasispin angular momentum is denoted $Q$.}
\label{ta-b}
\begin{tabular}{lcc|lcc|lcc|lcc}\hline
No & Term & 2Q &
No & Term & 2Q &
No & Term & 2Q &
No & Term & 2Q \\ \hline
1  & $^{8}S^{0}$ & 0 &
35 & $^{4}I^{1}$ & 4 &
69 & $^{2}F^{8}$ & 0 &
103& $^{2}K^{4}$ & 2 \\
2  & $^{6}P^{0}$ & 2 &
36 & $^{4}I^{2}$ & 2 &
70 & $^{2}F^{9}$ & 0 &
104& $^{2}K^{5}$ & 2 \\
3  & $^{6}D^{0}$ & 0 &
37 & $^{4}I^{3}$ & 2 &
71 & $^{2}F^{A}$ & 0 &
105& $^{2}K^{6}$ & 0 \\
4  & $^{6}F^{0}$ & 2 &
38 & $^{4}I^{4}$ & 0 &
72 & $^{2}G^{1}$ & 4 &
106& $^{2}K^{7}$ & 0 \\
5  & $^{6}G^{0}$ & 0 &
39 & $^{4}I^{5}$ & 0 &
73 & $^{2}G^{2}$ & 1 &
107& $^{2}L^{1}$ & 4 \\
6  & $^{6}H^{0}$ & 2 &
40 & $^{4}K^{1}$ & 2 &
74 & $^{2}G^{3}$ & 2 &
108& $^{2}L^{2}$ & 2 \\
7  & $^{6}I^{0}$ & 0 &
41 & $^{4}K^{2}$ & 2 &
75 & $^{2}G^{4}$ & 2 &
109& $^{2}L^{3}$ & 2 \\
8  & $^{4}S^{1}$ & 4 &
42 & $^{4}K^{3}$ & 0 &
76 & $^{2}G^{5}$ & 2 &
110& $^{2}L^{4}$ & 0 \\
9  & $^{4}S^{2}$ & 0 &
43 & $^{4}L^{1}$ & 2 &
77 & $^{2}G^{6}$ & 2 &
111& $^{2}L^{5}$ & 0 \\
10 & $^{4}P^{1}$ & 2 &
44 & $^{4}L^{2}$ & 0 &
78 & $^{2}G^{7}$ & 0 &
112& $^{2}M^{1}$ & 2 \\
11 & $^{4}P^{2}$ & 2 &
45 & $^{4}L^{3}$ & 0 &
79 & $^{2}G^{8}$ & 0 &
113& $^{2}M^{2}$ & 2 \\
12 & $^{4}D^{1}$ & 4 &
46 & $^{4}M^{0}$ & 2 &
80 & $^{2}G^{9}$ & 0 &
114& $^{2}M^{3}$ & 0 \\
13 & $^{4}D^{2}$ & 2 &
47 & $^{4}N^{0}$ & 0 &
81 & $^{2}G^{A}$ & 0 &
115& $^{2}M^{4}$ & 0 \\
14 & $^{4}D^{3}$ & 2 &
48 & $^{2}S^{1}$ & 0 &
82 & $^{2}H^{1}$ & 4 &
116& $^{2}N^{1}$ & 2 \\
15 & $^{4}D^{4}$ & 0 &
49 & $^{2}S^{2}$ & 0 &
83 & $^{2}H^{2}$ & 4 &
117& $^{2}N^{2}$ & 0 \\
16 & $^{4}D^{5}$ & 0 &
50 & $^{2}P^{1}$ & 4 &
84 & $^{2}H^{3}$ & 2 &
118& $^{2}O^{0}$ & 2 \\
17 & $^{4}D^{6}$ & 0 &
51 & $^{2}P^{2}$ & 2 &
85 & $^{2}H^{4}$ & 2 &
119& $^{2}Q^{0}$ & 0 \\
18 & $^{4}F^{1}$ & 4 &
52 & $^{2}P^{3}$ & 2 &
86 & $^{2}H^{5}$ & 2 &
120& $^{7}F^{0}$ & 1 \\
19 & $^{4}F^{2}$ & 2 &
53 & $^{2}P^{4}$ & 2 &
87 & $^{2}H^{6}$ & 2 &
121& $^{5}D^{1}$ & 3 \\
20 & $^{4}F^{3}$ & 2 &
54 & $^{2}P^{5}$ & 0 &
88 & $^{2}H^{7}$ & 2 &
122& $^{5}D^{2}$ & 1 \\
21 & $^{4}F^{4}$ & 2 &
55 & $^{2}D^{1}$ & 4 &
89 & $^{2}H^{8}$ & 0 &
123& $^{5}D^{3}$ & 1 \\
22 & $^{4}F^{5}$ & 0 &
56 & $^{2}D^{2}$ & 4 &
90 & $^{2}H^{9}$ & 0 &
124& $^{5}F^{1}$ & 3 \\
23 & $^{4}G^{1}$ & 4 &
57 & $^{2}D^{3}$ & 2 &
91 & $^{2}I^{1}$ & 4 &
125& $^{5}F^{2}$ & 1 \\
24 & $^{4}G^{2}$ & 2 &
58 & $^{2}D^{4}$ & 2 &
92 & $^{2}I^{2}$ & 2 &
126& $^{5}G^{1}$ & 3 \\
25 & $^{4}G^{3}$ & 2 &
59 & $^{2}D^{5}$ & 2 &
93 & $^{2}I^{3}$ & 2 &
127& $^{5}G^{2}$ & 1 \\
26 & $^{4}G^{4}$ & 2 &
60 & $^{2}D^{6}$ & 0 &
94 & $^{2}I^{4}$ & 2 &
128& $^{5}G^{3}$ & 1 \\
27 & $^{4}G^{5}$ & 0 &
61 & $^{2}D^{7}$ & 0 &
95 & $^{2}I^{5}$ & 2 &
129& $^{5}P^{0}$ & 1 \\
28 & $^{4}G^{6}$ & 0 &
62 & $^{2}F^{1}$ & 6 &
96 & $^{2}I^{6}$ & 0 &
130& $^{5}H^{1}$ & 1 \\
29 & $^{4}G^{7}$ & 0 &
63 & $^{2}F^{2}$ & 4 &
97 & $^{2}I^{7}$ & 0 &
131& $^{5}H^{2}$ & 1 \\
30 & $^{4}H^{1}$ & 2 &
64 & $^{2}F^{3}$ & 2 &
98 & $^{2}I^{8}$ & 0 &
132& $^{5}S^{0}$ & 3 \\
31 & $^{4}H^{2}$ & 2 &
65 & $^{2}F^{4}$ & 2 &
99 & $^{2}I^{9}$ & 0 &
133& $^{5}I^{1}$ & 3 \\
32 & $^{4}H^{3}$ & 2 &
66 & $^{2}F^{5}$ & 2 &
100& $^{2}K^{1}$ & 4 &
134& $^{5}I^{2}$ & 1 \\
33 & $^{4}H^{4}$ & 0 &
67 & $^{2}F^{6}$ & 2 &
101& $^{2}K^{2}$ & 2 &
135& $^{5}K^{0}$ & 1 \\
34 & $^{4}H^{5}$ & 0 &
68 & $^{2}F^{7}$ & 2 &
102& $^{2}K^{3}$ & 2 &
136& $^{5}L^{0}$ & 1 \\ \hline
\end{tabular}
\end{center}
\end{table}

\begin{table}
\vspace {0.5in} Table 2 (continued) \vspace {0.1in}
\begin{center}
\begin{tabular}{lcc|lcc|lcc}
\hline
No & Term & 2Q &
No & Term & 2Q &
No & Term & 2Q \\ \hline
137& $^{3}F^{1}$ & 5 &
171& $^{3}H^{8}$ & 1 &
205& $^{1}D^{6}$ & 1 \\
138& $^{3}F^{2}$ & 3 &
172& $^{3}H^{9}$ & 1 &
206& $^{1}G^{6}$ & 1 \\
139& $^{3}F^{6}$ & 1 &
173& $^{3}I^{1}$ & 3 &
207& $^{1}G^{7}$ & 1 \\
140& $^{3}F^{8}$ & 1 &
174& $^{3}I^{2}$ & 3 &
208& $^{1}G^{8}$ & 1 \\
141& $^{3}D^{1}$ & 3 &
175& $^{3}I^{3}$ & 1 &
209& $^{1}D^{4}$ & 3 \\
142& $^{3}D^{2}$ & 3 &
176& $^{3}I^{4}$ & 1 &
210& $^{1}G^{4}$ & 3 \\
143& $^{3}D^{3}$ & 1 &
177& $^{3}I^{5}$ & 1 &
211& $^{1}H^{1}$ & 3 \\
144& $^{3}D^{4}$ & 1 &
178& $^{3}I^{6}$ & 1 &
212& $^{1}H^{2}$ & 3 \\
145& $^{3}F^{3}$ & 3 &
179& $^{3}K^{1}$ & 3 &
213& $^{1}P^{0}$ & 1 \\
146& $^{3}F^{5}$ & 1 &
180& $^{3}K^{2}$ & 3 &
214& $^{1}H^{3}$ & 1 \\
147& $^{3}G^{1}$ & 3 &
181& $^{3}K^{3}$ & 1 &
215& $^{1}H^{4}$ & 1 \\
148& $^{3}G^{2}$ & 3 &
182& $^{3}K^{4}$ & 1 &
216& $^{1}S^{1}$ & 7 \\
149& $^{3}G^{4}$ & 1 &
183& $^{3}K^{5}$ & 1 &
217& $^{1}I^{1}$ & 5 \\
150& $^{3}G^{5}$ & 1 &
184& $^{3}K^{6}$ & 1 &
218& $^{1}S^{2}$ & 3 \\
151& $^{3}D^{5}$ & 1 &
185& $^{3}L^{1}$ & 3 &
219& $^{1}I^{2}$ & 3 \\
152& $^{3}F^{4}$ & 3 &
186& $^{3}L^{2}$ & 1 &
220& $^{1}I^{3}$ & 3 \\
153& $^{3}F^{7}$ & 1 &
187& $^{3}L^{3}$ & 1 &
221& $^{1}S^{3}$ & 1 \\
154& $^{3}F^{9}$ & 1 &
188& $^{3}M^{1}$ & 3 &
222& $^{1}I^{4}$ & 1 \\
155& $^{3}G^{3}$ & 3 &
189& $^{3}M^{2}$ & 1 &
223& $^{1}I^{5}$ & 1 \\
156& $^{3}G^{6}$ & 1 &
190& $^{3}M^{3}$ & 1 &
224& $^{1}S^{4}$ & 1 \\
157& $^{3}G^{7}$ & 1 &
191& $^{3}N^{0}$ & 1 &
225& $^{1}I^{6}$ & 1 \\
158& $^{3}P^{1}$ & 5 &
192& $^{3}O^{0}$ & 1 &
226& $^{1}I^{7}$ & 1 \\
159& $^{3}P^{2}$ & 3 &
193& $^{1}F^{2}$ & 1 &
227& $^{1}K^{1}$ & 3 \\
160& $^{3}P^{3}$ & 3 &
194& $^{1}F^{3}$ & 1 &
228& $^{1}K^{2}$ & 1 \\
161& $^{3}H^{1}$ & 5 &
195& $^{1}F^{4}$ & 1 &
229& $^{1}K^{3}$ & 1 \\
162& $^{3}H^{2}$ & 3 &
196& $^{1}D^{1}$ & 5 &
230& $^{1}L^{1}$ & 3 \\
163& $^{3}H^{3}$ & 3 &
197& $^{1}D^{2}$ & 3 &
231& $^{1}L^{2}$ & 3 \\
164& $^{3}H^{4}$ & 3 &
198& $^{1}D^{3}$ & 3 &
232& $^{1}L^{3}$ & 1 \\
165& $^{3}P^{4}$ & 1 &
199& $^{1}F^{1}$ & 3 &
233& $^{1}L^{4}$ & 1 \\
166& $^{3}H^{5}$ & 1 &
200& $^{1}G^{1}$ & 5 &
234& $^{1}M^{1}$ & 1 \\
167& $^{3}H^{6}$ & 1 &
201& $^{1}G^{2}$ & 3 &
235& $^{1}M^{2}$ & 1 \\
168& $^{3}P^{5}$ & 1 &
202& $^{1}G^{3}$ & 3 &
236& $^{1}N^{1}$ & 3 \\
169& $^{3}P^{6}$ & 1 &
203& $^{1}D^{5}$ & 1 &
237& $^{1}N^{2}$ & 1 \\
170& $^{3}H^{7}$ & 1 &
204& $^{1}G^{5}$ & 1 &
238& $^{1}Q^{0}$ & 1 \\ \hline

\end{tabular}
\end{center}
\end{table}

Single shell data are stored in the two arrays I and
B. The former consists of
\begin{itemize}
\item I(1) is the state number of the shell (see Tables \ref{ta-a}, \ref{ta-b}).
\item I(2) is the principal quantum number $n$.
\item I(3) is the orbital quantum number $l$.
\item I(4) is the number of  electrons in the subshell.
\item I(5) is the shell total angular momentum $L$ multiplied by two.
\item I(6) is the shell total angular momentum $S$ multiplied by two.
\item I(7) is the shell total quasispin $Q$ multiplied by two.
\end{itemize}
The array B contains
\begin{itemize}
\item B(1) is the shell quasispin $Q$.
\item B(2) is the shell total angular momentum $S$.
\item B(3) is the shell quasispin projection  $M_Q$.
\end{itemize}

These arrays are placed in Common blocks /TRK/ and /TRK2/.
In particular, these are:

\begin{tabbing}

\=Name~~~~~~~~~~\=Dimension~~~~\=Function\\
  \\
\>\bf{/TRK/}\>           \>\bf{The data of the orbitals for first two shells}\\
\>BDS1\>3\>the array B for the first shell of the ket function\\
\>BDS2\>3\>the array B for the second shell of the ket function\\
\>BKS1\>3\>the array B for the first shell of the bra function\\
\>BKS2\>3\>the array B for the second shell of the bra function\\
\>IBDS1\>7\>the array I for the first shell of the ket function\\
\>IBDS2\>7\>the array I for the second  shell of the  ket function\\
\>IBKS1\>7\>the array I for the first shell of the bra function\\
\>IBKS2\>7\>the array I for the second  shell of the bra function\\
  \\
\>\bf{/TRK2/}\>             \>\bf{The data of the orbitals for last two shells}\\
\>BDS3\>3\>the array B for the third shell of the ket function\\
\>BDS4\>3\>the array B for the fourth shell of the ket function\\
\>BKS3\>3\>the array B for the third shell of the bra function\\
\>BKS4\>3\>the array B for the fourth shell of the bra function\\
\>IBDS3\>7\>the array I for the third shell of the ket function\\
\>IBDS4\>7\>the array I for the fourth  shell of the ket function\\
\>IBKS3\>7\>the array I for the third shell of the bra function\\
\>IBKS4\>7\>the array I for the fourth  shell of the bra function\\
\end{tabbing}

In this library, the common blocks /MEDEFN/ and /FACT/ from the earlier
version of the MCHF atomic--structure package \cite{Fbook,F}, and the newly created
auxiliary COMMON blocks /KAMPAS/, /RIBOLS/, /RIBOLSF/, /RIBOF/, /RIBOLS3/,
are used.

\medskip

The authors would like to draw attention at the ordering of terms in
Table~\ref{ta-b}, which is tuned to simplifying the placement of tables of
$f$-- shell reduced coefficients of fractional parentage into DATA blocks.

\subsubsection{Subroutines}

\subsubsection*{The subroutine NUMTER}

FUNCTION NUMTER has several modes of operation, but only one of these is
possible while using it in other programs. If the input values are the
shell's total quasispin $Q$ multiplied by two (input argument I2Q),
the shell's total spin $S$ multiplied by two (input argument I2S),
the shell's total angular momentum $L$ multiplied by two
(input argument I2L) and the quantum number $l$, it will find the number
of the $s$--, $p$-- $d$-- shell's term, as numbered in Table \ref{ta-a}. The
other arguments of this subroutine, $NK$ and $ND$, should be set to 3.
In this mode, a COMMON block /MT/, is needed by the program, which is
defined in BLOCK DATA TERMLS.

\subsubsection*{The subroutine RUMT}

The subroutine RUMT has several modes of operation, but only one of these
is possible while using it in other programs. For an input of the orbital quantum
number $l$ (input argument $LL$) and the term number from
Table \ref{ta-a} (input argument $KNT$), it finds
the shell's total quasispin $Q$ multiplied by two (output argument LQ),
the shell's total spin $S$ multiplied by two (output argument LS) and
the shell's total angular momentum $L$ multiplied by two
(output argument L). It finds these characteristics for $s$--, $p$-- and $
d$--
shells. A COMMON blocks /MT/ is needed by this program, which is defined in
BLOCK DATA TERMLS.

\subsubsection*{The subroutine C0T5S}

This subroutine determines the value of the Clebsch--Gordan coefficients:
\begin{equation}
\label{eq:c0t5s}
\left[
\begin{array}{ccc}
Q   & \frac{1}{2} & C \\
QM & SM & CM
\end{array}
\right].
\end{equation}
The subroutine has the input arguments $Q$, $QM$, $SM$, $C$, $CM$ and
output argument $A$.
The subroutine performs its calculations by employing analytical
expressions from  Varshalovich {\it et al}~\cite{VMK}.

\subsubsection*{The subroutine C1E0SM}

This routine determines the value of the Clebsch--Gordan coefficients:
\begin{equation}
\label{eq:c1e0sm}
\left[
\begin{array}{ccc}
Q   & 1 & C \\
QM  & 0 & CM
\end{array}
\right].
\end{equation}
The subroutine has the input arguments $Q$, $QM$, $C$, $CM$ and
output argument $A$.
The subroutine performs its calculations by employing analytical
expressions from  Varshalovich {\it et al}~\cite{VMK}.

\subsubsection*{The subroutine C1E1SM}

This subroutine determines the value of the Clebsch--Gordan coefficients:
\begin{equation}
\label{eq:c1e1sm}
\left[
\begin{array}{ccc}
Q   & 1 & C \\
QM  & 1 & CM
\end{array}
\right].
\end{equation}
The subroutine has the input arguments $Q$, $QM$, $C$, $CM$ and
output argument $A$.
The subroutine performs its calculations by employing analytical
expressions from  Varshalovich {\it et al}~\cite{VMK}.

\subsubsection*{The subroutine SIXJ}

This routine determines the value of the $6j$--coefficients:
\begin{equation}
\label{eq:sixj}
\left\{
\begin{array}{ccc}
I/2 & J/2 & K/2 \\
L/2 & M/2 & N/2
\end{array}
\right\}.
\end{equation}
The subroutine has the input arguments $I$, $J$, $K$, $L$, $M$, $N$, $ITIK$ and
output argument $SI$. If the parameter $ITIK=0$, the subroutine does not
check the triangular conditions for $6j$--coefficient. In other cases it
checks these. If any of the parameters of $6j$--coefficient is equal to
$0$, $\frac{1}{2}$, $1$, $\frac{3}{2}$, $2$, $3$, $4$, the subroutine
calculates the $6j$--coefficients according to analytical formulas \cite{VMK,JS}.
Otherwise, the customary calculations are performed. In that case, the
COMMON block /FACT/ must be defined. This is done by addressing the
SUBROUTINE FACTRL from the library {\bf MCHF\_LIB\_COM} \cite{Fbook,F}.

\subsubsection*{The subroutine NINE}

\begin{equation}
\label{eq:nine}
\left\{
\begin{array}{ccc}
J1/2 & J2/2 & J3/2 \\
L1/2 & L2/2 & L3/2 \\
K1/2 & K2/2 & K3/2
\end{array}
\right\}.
\end{equation}
The subroutine has the input arguments $J1$, $J2$, $J3$, $L1$, $L2$, $L3$,
$K1$, $K2$, $K3$, $I$ and output arguments $IN$ and $AA$.
If the parameter $I=1$, the subroutine only checks the triangular conditions
of a $9j$-- coefficient. If these are not satisfied, then $IN=0$, and $IN=1$
otherwise. At other values of $I$, the subroutine calculates the value of
$9j$-- coefficient and assigns it to the output parameter $AA$.

\subsubsection*{The subroutine SLS}

This subroutine determines the value of the RCFP:
\begin{equation}
\label{eq:sq-a}
(\ l\ QLS\ |||\ a^{(qls)}\ |||\ l\ Q'L'S'\ )
\end{equation}
for $p$--, $d$--, and $f$-- shells (see (39) in \cite{method2}).
The routine uses the table of reduced matrix elements of the $a^{(qls)}$
tensor operator from \cite{method5}).
The subroutine has the following arguments:
\begin{enumerate}
\item L is the orbital quantum number $l$.
\item IT is the state number of the bra function
(see Tables \ref{ta-a}, \ref{ta-b}).
\item LQ is the quasispin $Q$ for the bra function multiplied by two.
\item LL is the total angular momentum $L$ for the bra function multiplied by
      two.
\item LS is the total angular momentum $S$ for the bra function multiplied by
      two.
\item ITS is the state number of the ket function.
\item LQS is the quasispin $Q$ for the ket function multiplied by two.
\item LLS is the total angular momentum $L$ for the ket function multiplied by
      two.
\item LSS is the total angular momentum $S$ for the ket function multiplied by
      two.
\item
S is the value of the reduced matrix element (\ref{eq:sq-a}) which is
      returned by the subroutine.
\end{enumerate}

\subsubsection*{The subroutine RWLS}

The routine determines the value of the reduced matrix element:
\begin{equation}
\label{eq:sq-aa}
(\ l\ QLS\ |||\ [~a^{(qls)} \times a^{(qls)}~]^{(k_{1}k_{2}k_{3})}\ |||\ l\ Q'L'S'\ ).
\end{equation}
The routine uses the tables of reduced matrix elements of the tensor
operator $[a^{(qls)} \times a^{(qls)}]^{(k_{1}k_{2}k_{3})}$
for $s$--, $p$-- and $d$-- subshells (see \v Spakauskas {\it et al} ~\cite{SKR}),
and for the $f$-- shell the expression (34) from paper \cite{method1} is used.
The subroutine does not calculate the simple case of $k_{1}=k_{2}=k_{3}=0$,
because then the operator is just
$[~a^{(qls)} \times a^{(qls)}~]^{(000)}=-(2l+1)^{1/2}$
(expression (15.54) in Rudzikas~\cite{R}).
The subroutine has the formal arguments:

\begin{enumerate}
\item K1 is the rank $k_{1}$.
\item K2 is the rank $k_{2}$.
\item K3 is the rank $k_{3}$.
\item L is the orbital quantum number $l$.
\item J1 is the state number of the bra function (see Tables \ref{ta-a}, \ref{ta-b}).
\item J2 is the state number of the ket function.
\item W is the value of the reduced matrix element (\ref{eq:sq-aa}) which is returned by
the subroutine.
\end{enumerate}

\subsubsection*{The subroutine W1}
\label{sec-w1}

This subroutine determines the value of the matrix element:
\begin{equation}
\label{eq:sq-b}
(\ l^{N}\ QLS\ ||\ [a^{(qls)}_{m_{q1}} \times a^{(qls)}_{m_{q2}}]^{(k_{l}k_{s})} ||\ l^{N'}\ Q'L'S'\ ).
\end{equation}
While calculating cases where the orbital number $l$=0, 1, 2, 3 and the shell
occupation number $N > 2$, the program relies on the expression
(31) from the paper \cite{method1}. In that case,
the subroutine finds the Clebsch--Gordan coefficient which gives the 
dependence
on the shell occupation number. If the tensor product (\ref{eq:sq-b}) consists
of either two electron creation or two annihilation operators then C1E1SM is
called. Otherwise CLE0SM is called. The subroutine RWLS finds the reduced matrix
elements of the operator $[~a^{(qls)} \times a^{(qls)}~]^{(k_{l}k_{s})}$.
In other cases, the program calculates according to the expression (40)
from paper \cite{method2}.
The subroutine has the formal arguments:
\begin{enumerate}

\item IK is the array I for the bra function.
\item BK is the array B for the bra function.
\item ID is the array I for the ket function.
\item BD is the array B for the ket function.
\item K2 is the rank $k_{l}$.
\item K3 is the rank $k_{s}$.
\item  QM1, QM2 are the quasispin projections in (\ref{eq:sq-b}).
\item W is the value of the reduced matrix element (\ref{eq:sq-b}) which is returned by the subroutine.
\end{enumerate}

\subsubsection*{The subroutine AWP1LS}

The routine determines the value of the matrix elements:

\begin{equation}
\label{eq:sq-c}
(\ l^{N}\ QLS\ ||\ [ a^{(qls)}_{m_{q1}} \times [\ a^{(qls)}_{m_{q2}} \times a^{(qls)}_{m_{q3}}\ ]^{(k_{1}k_{2})} ]^{(k_{l}k_{s})}\ ||\ l^{N'}\ Q'L'S'\ ).
\end{equation}
While calculating cases where the orbital number $l$=0, 1, 2, 3 and the
shell's occupation number $N > 2$, the program relies on the expression (31)
from the paper \cite{method1}. In that case,
the subroutine IZAS1 checks that the subshell has a state
with the specified characteristics. The subroutine ITLS2 finds the first and
the last numbers of the state from the running intermediate sum in array
MT.  RUMT finds the shell's total angular momentum $LS$ and quasispin  $Q$ for
each intermediate state. The routine IXJTIK checks all triads.
The subroutine C0T5S finds the Clebsch--Gordan coefficient giving the
dependence on the shell's occupation number and SLS finds the reduced matrix
element of $a^{(qls)}$ tensor operator
(see in Section~\ref{sec-sls}). The second part of
the expression is calculated by the routine W1
(see in Section~\ref{sec-w1}). The routine SIXJ finds $6j$--symbol.
In other cases, the program calculates according to the
expression (40) from paper \cite{method2}.
The subroutine has the arguments:
\begin{enumerate}

\item IK is the array I for the bra function.
\item BK is the array B for the bra function.
\item ID is the array I for the ket function.
\item BD is the array B for the ket function.
\item K1 is the rank $k_{1}$.
\item K2 is the rank $k_{2}$.
\item K3 is the rank $k_{l}$.
\item BK4 is the rank $k_{s}$.
\item  QM1, QM2 and QM3 are the quasispin projections in (\ref{eq:sq-c}).
\item AW is the value of the reduced matrix element (\ref{eq:sq-c}) which is
returned by the subroutine.
\end{enumerate}

\subsubsection*{The subroutine WAP1LS}

This subroutine determines the value of the matrix elements:
\begin{equation}
\label{eq:sq-d}
(\ l^{N}\ QLS\ ||\ [ [\ a^{(qls)}_{m_{q1}} \times a^{(qls)}_{m_{q2}}\ ]^{(k_{1}k_{2})} \times a^{(qls)}_{m_{q3}}\ ]^{(k_{l}k_{s})}\ ||\ l^{N'}\ Q'L'S'\ ).
\end{equation}
The structure of the routine WAP1LS is the same as that of AWP1LS.
The subroutine has the formal arguments:
\begin{enumerate}

\item IK is the array I for the bra function.
\item BK is the array B for the bra function.
\item ID is the array I for the ket function.
\item BD is the array B for the ket function.
\item K1 is the rank $k_{1}$.
\item K2 is the rank $k_{2}$.
\item K3 is the rank $k_{l}$.
\item BK4 is the rank $k_{s}$.
\item  QM1, QM2 and QM3 are the quasispin projections in (\ref{eq:sq-d}).
\item WA is the value of the reduced matrix element (\ref{eq:sq-d}) which is returned 
by the subroutine.
\end{enumerate}

\subsubsection*{The subroutine WWLS1}

The routine determines the value of matrix elements:

\begin{equation}
\label{eq:sq-e}
(\ l^{N}\ QLS\ ||\ [\ [\ a^{(qls)}_{m_{q1}} \times a^{(qls)}_{m_{q2}}\ ]^{(k_{l}k_{s})} \times [\ a^{(qls)}_{m_{q3}} \times a^{(qls)}_{m_{q4}}\ ]^{(k_{l}k_{s})} ]^{(00)}\ ||\ l^{N'}\ QLS\ ).
\end{equation}
The routine WWLS1 uses the routines ITLS, RUMT, IZAS1 for calculation of this sort
of the matrix element. The subroutine W1 calculates the first and the second
parts of the operator calculates.
The subroutine has the formal arguments:
\begin{enumerate}

\item IK is the array I for the bra function.
\item BK is the array B for the bra function.
\item ID is the array I for the ket function.
\item BD is the array B for the ket function.
\item K2 is the rank $k_{l}$.
\item K3 is the rank $k_{s}$.
\item  QM1, QM2, QM3 and QM4 are the quasispin projections in (\ref{eq:sq-e}).
\item WW is the value of the reduced matrix element (\ref{eq:sq-e}) which is returned by the subroutine.
\end{enumerate}

\subsubsection*{The subroutine WWPLS1}

The routine determines the value of matrix elements:

\begin{equation}
\label{eq:sq-ep}
(\ l^{N}\ QLS\ ||\ [\ [\ a^{(qls)}_{m_{q1}} \times a^{(qls)}_{m_{q2}}\ ]^{(k_{1}k_{2})} \times [\ a^{(qls)}_{m_{q3}} \times a^{(qls)}_{m_{q4}}\ ]^{(k_{3}k_{4})} ]^{(k_{l}k_{s})}\ ||\ l^{N'}\ QLS\ ).
\end{equation}
The routine WWPLS1 uses the routines ITLS, RUMT, IZAS1 for calculation of this sort
of the matrix element. The subroutine W1 calculates the first and the second
parts of the operator calculates. The routine SIXJ finds $6j$--symbol.
The subroutine uses the COMMON block /TRK/, which contains arrays I and B
for the bra and ket function (ID1, IK1,BD1 and BK1).
It uses the expression (40) from \cite{method2}.
The subroutine has the formal arguments:
\begin{enumerate}

\item K1 is the rank $k_{1}$.
\item K2 is the rank $k_{2}$.
\item K3 is the rank $k_{3}$.
\item K4 is the rank $k_{4}$.
\item K5 is the rank $k_{l}$.
\item K6 is the rank $k_{s}$.
\item  QM1, QM2, QM3 and QM4 are the quasispin projections in (\ref{eq:sq-ep}).
\item WW is the value of the reduced matrix element (\ref{eq:sq-ep}) which is returned by the subroutine.
\end{enumerate}

\subsection{SAI\_SQLS2}
\label{sec-sls2}

The library {\bf SAI\_SQLS2} is the second part of Standard quantities in $LS$-- coupling.
Everything in it is related to the $f$-- shells. The tables of CFP
for $f$-- shell (see \cite{method5}) and the term characteristics
are in this library. This library can be used in other programs to its full
extent, by employing programs from {\bf SAI\_SQLS1} library. These are the
subroutines SLS, RWLS, W1, AWP1LS, WAP1LS, WWLS1, WWPLS1. The remaining two
subroutines may also be used independently.

\subsubsection*{The subroutine NUMTERF}

The subroutine has several modes of operation. But only one of these is
useful while operated independently from the MCHF atomic structure package
\cite{Fbook,F}.
In that case, after giving number $Nr$ from Table \ref{ta-b}
(input argument I2N), the values of
the shell total spin $S$ multiplied by two (input arguments I2S and N),
the shell total angular momentum $L$ multiplied by two,
(input argument I2L) and
the shell total quasispin $Q$ multiplied by two (input argument I2Q)
for the FUNCTION NUMBER, it finds the number of $f$-- shell term, as
numbered in Table \ref{ta-b}. The subroutine needs the COMMON block /MT67/,
which is defined by BLOCK DATA TERMF.

\subsubsection*{The subroutine RUMT67}

For an input of the term member from Table \ref{ta-b} (input argument $KNT$),
the subroutine RUMT67 finds the number $Nr$ from
Table \ref{ta-b} (output argument NR),
the shell total quasispin $Q$ multiplied by two (output argument LQ),
the shell total spin $S$ multiplied by two (output argument LS) and
the shell total angular momentum $L$ multiplied by two
(output argument L). The subroutine needs the COMMON block /MT67/ which
is defined by BLOCK DATA TERMF.

\section{Examples}

A driver program (see Figure 1) accompanying the library SQLSF
illustrates three examples.

\medskip

\begin{figure}
\begin{scriptsize}
\begin{verbatim}
      IF(ICASE .EQ. 1) THEN
        WRITE(6,'(/A)') ' orbital quantum number l (I1) '
        READ(5,'(I1)') L
        IF(L.LT.3) THEN
          WRITE(6,'(/A)') ' 2*Q 2*L 2*S for bra function (3I2) '
          READ(5,'(3I2)') IQB,ILB,ISB
          JB=NUMTER(IQB,ISB,ILB,L,3,3)
          WRITE(6,'(/A)') ' 2*Q 2*L 2*S for ket function (3I2) '
          READ(5,'(3I2)') IQK,ILK,ISK
          JK=NUMTER(IQK,ISK,ILK,L,3,3)
        ELSEIF(L.EQ.3) THEN
          WRITE(6,'(/A)') ' 2*Q 2*L 2*S Nr for bra function (4I2) '
          READ(5,'(4I2)') IQB,ILB,ISB,INB
          JB=NUMTERF(INB,ISB,ILB,ISB,IQB)
          WRITE(6,'(/A)') ' 2*Q 2*L 2*S Nr for ket function (4I2) '
          READ(5,'(4I2)') IQK,ILK,ISK,INK
          JK=NUMTERF(INK,ISK,ILK,ISK,IQK)
        ENDIF
        CALL SLS(L,JB,IQB,ILB,ISB,JK,IQK,ILK,ISK,S)
        WRITE(6,'(/A,F17.7)') ' Value= ',S
      ELSEIF(ICASE .EQ. 2) THEN
        WRITE(6,'(/A)') ' orbital quantum number l (I1) '
        READ(5,'(I1)') L
        WRITE(6,'(/A)') ' number electrons in the shell N (I1) '
        READ(5,'(I1)') N 
        WRITE(6,'(/A)') ' bra and ket number of term (2I3) '
        READ(5,'(2I3)') JB,JK
        IF(L.LT.3) THEN
          CALL RUMT(JB,L,IQB,ISB,ILB)
          CALL RUMT(JK,L,IQK,ISK,ILK)
        ELSEIF(L.EQ.3) THEN
          CALL RUMT67(JB,NRB,IQB,ISB,ILB)
          CALL RUMT67(JK,NRK,IQK,ISK,ILK)
        ENDIF
        CALL SLS(L,JB,IQB,ILB,ISB,JK,IQK,ILK,ISK,S)
        QK=DBLE(IQK)*0.5
        QB=DBLE(IQB)*0.5
        QMK=DBLE(N-1-2*L-1)*0.5
        QMB=DBLE(N-2*L-1)*0.5
        QM=0.5
        CALL C0T5S(QK,QMK,QM,QB,QMB,A)
        CFP=S*A/DSQRT(DBLE(N*(IQB+1)*(ILB+1)*(ISB+1)))
        IF(((N-1)/2)*2.NE.N-1) CFP=-CFP
        WRITE(6,'(/A,F17.7)') ' Value= ',CFP
      ELSEIF(ICASE .EQ. 3) THEN
        WRITE(6,'(/A)') ' orbital quantum number l (I1) '
        READ(5,'(I1)') L
        WRITE(6,'(/A)') ' number electrons in the shell N (I1) '
        READ(5,'(I1)') N 
        WRITE(6,'(/A)') ' first and last number of term for bra (2I3) '
        READ(5,'(2I3)') JFB,JLB
        WRITE(6,'(/A)') ' first and last number of term for ket (2I3) '
        READ(5,'(2I3)') JFK,JLK
        WRITE(6,'(/A)') ' the ranks  K1 K2 K3 (3I1) '
        READ(5,'(3I1)') K1,K2,K3
        WRITE(6,'(3I1)') K1,K2,K3
        DO 2 JTB=JFB,JLB
          IF(L.LT.3) THEN
            CALL RUMT(JTB,L,IQB,ISB,ILB)
          ELSEIF(L.EQ.3) THEN
            CALL RUMT67(JTB,NRB,IQB,ISB,ILB)
          ENDIF
          DO 3 JTK=JFK,JLK
            IF(L.LT.3) THEN
              CALL RUMT(JTK,L,IQK,ISK,ILK)
            ELSEIF(L.EQ.3) THEN
              CALL RUMT67(JTK,NRK,IQK,ISK,ILK)
            ENDIF
            CALL RWLS(K1,K2,K3,L,JTB,JTK,W)
            QK=DBLE(IQK)*0.5
            QB=DBLE(IQB)*0.5
            QM=DBLE(N-2*L-1)*0.5
            CALL C1E0SM(QK,QM,QB,QM,A)
            VK=-0.5*W*A/DSQRT(DBLE((IQB+1)*(2*K2+1)))
            WRITE(6,'(/A,/2I3,F17.7)') 
     :      ' number for bra : number for ket : value ',JTB,JTK,VK
    3     CONTINUE
    2   CONTINUE
      ENDIF
\end{verbatim}
\end{scriptsize}
{\bf Figure 1:}
\hspace{0.2cm}
{\rm The program for three examples}
\end{figure}

The  examples show that some of the subroutines contained in the
libraries may serve as an electronic version of
Nielson and Koster \cite{NK} or \cite{method5} tables. As such
they may serve as a basis for extending the capabilities of
programs that rely on the principle of calculating the determinants
to arbitrarily filled $f$-- shells
(see for example Eissner {\it et al}~\cite{EJN} or Zatsarinny~\cite{Z}).

\medskip

Here we describe each example separately, the EXAMPLES RUN OUTPUT
being shown in Table 4.

\subsection*{Case 1. Finding the RCFP}

In this case the subroutine finds the value of a RCFP
$(\ f\ ^{8}S^{0}\ |||\ a^{(qls)}\ |||\ f\ ^{7}F^{0}\ )$.
From Table IV of \cite{method5} we see that the value of this coefficient is
equal to
$$(\ f\ ^{8}S^{0}\ |||\ a^{(qls)}\ |||\ f\ ^{7}F^{0}\ )=-4 \sqrt{7}.$$
This is in accordance with the result obtained with $SLS$ subroutine
(see in Section~\ref{sec-sls}).

\medskip

It is necessary to mention that in the calls to this and some other subroutine,
the input of a term number and term characteristics is needed. In this case the
user indicates only the term characteristics, and finds the term number by using
the subroutine NUMTER (see in Section~\ref{sec-sls}) or NUMTERF
(see in Section~\ref{sec-sls2}). In other examples we will present a way,
showing how to call subroutines of analogous type, by defining just the
term number by the user.

\subsection*{Case 2. Calculating the CFP}

This case illustrates the finding of a CFP
value, using the $SLS$ program. It may be useful for programs that are based
on the tables of CFP, which are much more
extended than the tables of RCFP.
The values of CFP are found by
programming the expression (\ref{eq:gi}).

\medskip

The value of a CFP $(\ f^{7}\ ^{6}D^{0}\ ||\ f^{6}\ ^{5}P^{0}\ f\ )$
is found in the example. We see from Nielson and Koster \cite{NK}
that the value of this CFP is equal to
$$(\ f^{7}\ ^{6}D^{0}\ ||\ f^{6}\ ^{5}P^{0}\ f\ )=-\frac{\sqrt{3}}{7}.$$
This is in accordance with the result obtained with $SLS$ subroutine.

\medskip

See \cite{method5,GFa} for details of the encoding of the $f$-- shell terms.
See \cite{method5} for a more extensive use of the tables of RCFP in
finding the CFP. In that paper, also the problems related to that task
are discussed and the ways to solve these are indicated.

\subsection*{Case 3. Calculating the matrix elements of $V^{k}$ operator}

In this section it is demonstrated how to calculate the $V^{11}$
operator,
using the subroutine RWLS (see in Section~\ref{sec-sls}), which calculates
the reduced matrix elements (\ref{eq:sq-aa}). For that purpose the relation
(16.34) from Rudzikas \cite{R} is used. While using the $U^{k}$ and $V^{k1}$
tables, one must pay attention to various phase conventions used in the
literature.
In addition, small differences in the definitions of $U^{k}$ occur.
Some authors, Karazija {\it et al}~\cite{KVRJ} among them, tabulate the
submatrix elements
\begin{equation}
\label{eq:ex-a}
(l^{N} \alpha SL || U^{k} || l^{N} \alpha ^{\prime}S^{\prime}L^{\prime}),
\end{equation}
others, like Nielson and Koster~\cite{NK} or Cowan~\cite{Cowan}, tabulate
\begin{equation}
\label{eq:ex-b}
(l^{N} \alpha L || U^{k} || l^{N} \alpha ^{\prime}L^{\prime}),
\end{equation}
although they use the notation of (\ref{eq:ex-a}). Meanwhile, the relation
between these two coefficients is:
\begin{equation}
\label{eq:ex-c}
(l^{N} \alpha SL || U^{k} || l^{N} \alpha ^{\prime}S^{\prime}L^{\prime})=
\delta(S,S ^{\prime}) \sqrt{(2S+1)}
(l^{N} \alpha L || U^{k} || l^{N} \alpha ^{\prime}L^{\prime}).
\end{equation}
The submatrix elements are defined as (\ref{eq:ex-a}) if we use the
relations between matrix elements of
$W^{(k_1k_2k_3)}$ and $U^{k}$  as presented in the
Rudzikas monograph~\cite{R}.
Reduced matrix elements of the operator $V^{11}$
$(f^{7}\ ^{6}P^{0} || V^{11} || f^{7}\ ^{4}S^{1})$ are calculated in the
example.
The numerical value of this reduced matrix element
is taken from the Nielson and Koster \cite{NK} tables
$$(f^{7}\ ^{6}P^{0} || V^{11} || f^{7}\ ^{4}S^{1})~=  -\sqrt{\frac{2}{7}}$$
and agree with our value.

\section*{Table 4. EXAMPLES RUN OUTPUT}

\subsection*{Case 1. Finding a reduced coefficient of fractional parentage}

\begin{scriptsize}
\begin{verbatim}
>> Example
            ==========================================
             E X A M P L E S   for  L I B R A R I E S
            ==========================================
 Which case? (1,2,3)
>1
 CASE 1
 ===========
 orbital quantum number l (I1)
>3
 2*Q 2*L 2*S Nr for bra function (4I2)
> 0 0 7 0
 2*Q 2*L 2*S Nr for ket function (4I2)
> 1 6 6 0
 bra (number l 2*Q 2*L 2*S)
          1  3  0   0   7
 ket (number l 2*Q 2*L 2*S)
        120  3  1   6   6
 Value=       -10.583005244258363
 END OF CASE 1
 ===========
\end{verbatim}
\end{scriptsize}
\subsection*{Case 2. Calculation of the coefficient of fractional parentage}
\begin{scriptsize}
\begin{verbatim}
>> Example
            ==========================================
             E X A M P L E S   for  L I B R A R I E S
            ==========================================
 Which case? (1,2,3)
>2
 CASE 2
 ===========
 orbital quantum number l (I1)
>3
 number electrons in the shell N (I1)
>7
 bra and ket number of term (2I3)
  3129
 bra (number l 2*Q 2*L 2*S  N)
          3  3  0   4   5   7
 ket (number l 2*Q 2*L 2*S)
        129  3  1   2   4
 Value=        -0.247435829652697
 END OF CASE 2
 ===========
\end{verbatim}
\end{scriptsize}
\subsection*{Case 3. Calculation of the matrix elements of operator $V^{k}$}
\begin{scriptsize}
\begin{verbatim}
>> Example
            ==========================================
             E X A M P L E S   for  L I B R A R I E S
            ==========================================
 Which case? (1,2,3)
>3
 CASE 3
 ===========
 orbital quantum number l (I1)
>3
 number electrons in the shell N (I1)
>7
 first and last number of term for bra (2I3)
>  2  2
 first and last number of term for ket (2I3)
>  8 8
 the ranks  K1 K2 K3 (3I1)
>111
111
 number for bra : number for ket : value
  2  8       -0.534522483824849
 END OF CASE 3
 ===========
\end{verbatim}
\end{scriptsize}

\section{Conclusion}

Accurate theoretical determination of atomic energy levels, orbitals and radiative
transition data requires the calculation of matrix elements of physical operators 
accounting for relativistic and correlation effects. 
The spin--angular integration of these matrix elements is typically based on
standard quantities like the matrix elements of the unit tensor, the (reduced)
coefficients of fractional parentage as well as a number of other reduced 
matrix elements concerning various products of electron creation and 
annihilation operators.
These quantities arise very frequently in
large--scale computations of
open--shell atoms using multiconfiguration Hartree--Fock or configuration interaction
approaches, in calculating the angular parts of
effective operators from many--body perturbation theory or for evaluating relativistic
hamiltonian in $LS$--coupling as well as in various versions of semiempyrical methods.
The library SQ is assigned for the calculation of all these standard quantities.
It can be used as "electronic tables" of standard quantities for
evaluating general matrix elements for $LS$--coupled functions, too.

\medskip

The library SQ is created involving the
angular methodology of \cite{method1,method2,method3,method5,method7,method6}.
For $LS$--coupling subshells states, the library provides coefficients and matrix
elements for all subshells (nl) with $l=0$, $1$, $2$ and $3$, and $l^{2}$ for $l \ge 3$.

\medskip

Program is obtainable from State Institute of Theoretical Physic and Astronomy,
A. Go\v stauto 12, Vilnius, 2600, LITHUANIA.~~ E-mail: gaigalas@itpa.lt.

\newpage

"Prongramin\. e biblioteka atom\c u strukt\= uros teorijos standartiniams
dyd\v ziams skai\v ciuoti."

\medskip

G. Gaigalas

\medskip
\medskip
\medskip

Santrauka

\medskip

Straipsnyje pateikta biblioteka (paprogrami\c u rinkinys), para\v syta
Fortran~77 programavimo kalba. Ji skirta standartini\c u dyd\v zi\c u
skaiciavimui, su kuriais susiduriame atomo teorijoje nagrin\. ejant
fizikini\c u bei efektyvini\c u operatori\c u matricinius elementus
$LS$ ry\v syje. Visu pirma, tai kilminiai koeficientai bei vienetini\c u 
tenzori\c u
$U^{k}$ ir $V^{k1}$ matriciniai elementai, kurie paprastai pasirodo
nagrin\. ejant matricinius elementus klasikiniais metodais. Be to 
biblioteka skirta ir toki\c u standartini\c u dyd\v zi\c u 
nagrin\. ejimui, kurie
pasirodo naudojant kvazisukinio formalizm\c a. Tai 
subkilminiai keoficientai, bei 
redukuotiniai
kvazisukinio erdv\. eje, vienetiniai tenzoriai
$W^{(k_q k_l k_s)}$. Taigi darbe pateikta biblioteka yra bendro 
pob\= ud\v zio ir skirta matricini\c u element\c u skai\v ciavimui 
bet kokia mokslin\. eje literat\= uroje \v zinoma metodika. 
Pati biblioteka yra para\v syta
remiantis antrinio kvantavimo suri\v stame tenzoriniame pavidale 
formalizmu,
judesio kiekio momento teorija trijose erdv\. ese (orbitiniame, 
sukininiame ir kvazisukininiame) ir grafine judesio kiekio momento 
teorija. \v Si biblioteka i\v sple\v cia atomo teorijos galimybes, 
kadangi
joje yra pilnai \c itraukti f sluoksniai su bet kokiu sluoksnio 
u\v zpildymo skai\v ciumi, bei leid\v zia pagreitinti jau esamas 
programas.

\end{document}